\documentclass[aps,onecolumn,superscriptaddress]{revtex4-2}

\usepackage[T1]{fontenc}
\usepackage[utf8]{inputenc}
\usepackage{lmodern}
\usepackage[english]{babel}
\usepackage{graphicx}
\usepackage{acronym}
\usepackage{dcolumn}
\usepackage{bm}
\usepackage[mathlines]{lineno}

\usepackage[table]{xcolor}
\usepackage{xintexpr}
\usepackage{pgfplotstable}
\usepackage{amssymb}
\usepackage[cmex10]{amsmath}
\usepackage{amsfonts}
\usepackage{dsfont}
\usepackage{fontawesome}
\usepackage{amsthm}
\usepackage{hyperref}
\usepackage{cleveref}
\usepackage{braket}
\usepackage{multirow}
\usepackage{ifthen}
\usepackage[nomessages]{fp}
\usepackage[normalem]{ulem}
\usepackage{enumerate}

\crefname{equation}{eq.}{eqs.}
\Crefname{equation}{Eq.}{Eqs.}

\usepackage{tikz}
\usetikzlibrary{positioning,calc,mindmap,fadings,shapes.arrows,shadows,fit,backgrounds}

\pgfplotsset{compat=1.16}

\newcommand{\ve}[1]{\ensuremath{\mathbf{#1}}} 
\newcommand{\R}{\ensuremath{\mathbb{R}}} 
\newcommand{\N}{\ensuremath{\mathbb{N}}} 
\newcommand{\E}{\ensuremath{\mathbb{E}}} 
\newcommand{\indicator}{\ensuremath{\mathds{1}}} 

\newcommand{\iid}{\textbf{IID}}
\newcommand{\nm}{\textbf{NM}}
\newcommand{\vs}{\textbf{VS}}
\newcommand{\svm}{\textbf{m-SVM-single}}
\newcommand{\mlp}{\textbf{m-MLP-single}}
\newcommand{\svmt}{\textbf{m-SVM}}
\newcommand{\mlpt}{\textbf{m-MLP}}
\newcommand{\gru}{\textbf{m-GRU}}
\newcommand{\lstm}{\textbf{m-LSTM}}
\newcommand{\bigru}{\textbf{m-biGRU}}
\newcommand{\bilstm}{\textbf{m-biLSTM}}
\newcommand{\bigrua}{\textbf{m-biGRU-att}}
\newcommand{\bilstma}{\textbf{m-biLSTM-att}}
\newcommand{\bigrum}{\textbf{m-biGRU-max}}
\newcommand{\bilstmm}{\textbf{m-biLSTM-max}}

\DeclareMathOperator*{\argmax}{arg\,max}
\DeclareMathOperator*{\argmin}{arg\,min}

\definecolor{neuronT}{gray}{1}
\definecolor{neuronB}{gray}{0.5}
\definecolor{merge}{rgb}{1,0.5,0.5}
\definecolor{noiseaT}{rgb}{1,1,0.6}
\definecolor{noisebT}{rgb}{0.9,0.8,1}
\definecolor{noiseaB}{rgb}{0.8,0.8,0.4}
\definecolor{noisebB}{rgb}{0.7,0.6,1}
\definecolor{graph1}{rgb}{1,0.6,0}
\definecolor{graph2}{rgb}{0.5,1,0.5}
\definecolor{graph3}{rgb}{0.4,0.4,1}
\definecolor{graph4}{rgb}{0.5,1,1}
\definecolor{plate1}{rgb}{1,1,0.8}
\definecolor{plate2}{rgb}{0.8,1,0.8}
\definecolor{plate1lab}{rgb}{1,0,0}
\definecolor{plate2lab}{rgb}{1,0,0}
\definecolor{arrowT}{rgb}{1,0.8,0.8}
\definecolor{arrowB}{rgb}{1,0.5,0.5}
\definecolor{layerT}{rgb}{0.8,1,0.8}
\definecolor{layerB}{rgb}{0.6,0.8,0.6}
\definecolor{parameterT}{rgb}{1,0.8,0.8}
\definecolor{parameterB}{rgb}{1,0.5,0.5}

\tikzfading[name=shadowfading, top color=transparent!0, bottom color=transparent!95]
\tikzstyle{myShadow}=[fill=black, shadow xshift=0.2ex, shadow yshift=-0.6ex, path fading=shadowfading, fading angle=45]

\tikzstyle{every neuron}=[circle, top color=neuronT, bottom color=neuronB, minimum size=0.4cm, general shadow=myShadow]
\tikzstyle{neuron missing}=[draw=none, fill=none, opacity=0, text opacity=1, scale=2,text height=0.333cm,execute at begin node=\color{black}$\vdots$]
\tikzstyle{noisea}=[top color=noiseaT, bottom color=noiseaB, minimum size=0.8cm, general shadow=myShadow, rounded corners=0.1cm]
\tikzstyle{noiseb}=[top color=noisebT, bottom color=noisebB, minimum size=0.8cm, general shadow=myShadow, rounded corners=0.1cm]
\tikzstyle{merge}=[circle, draw, fill=merge, general shadow=myShadow]
\tikzstyle{graph}=[concept, minimum size=0.5cm, general shadow=myShadow]
\tikzstyle{graph1}=[graph, concept color=graph1]
\tikzstyle{graph2}=[graph, concept color=graph2]
\tikzstyle{graph3}=[graph, concept color=graph3]
\tikzstyle{graph4}=[graph, concept color=graph4]
\tikzstyle{bigArrow}=[top color=arrowT, bottom color=arrowB, general shadow=myShadow, single arrow,minimum height=1cm, minimum width=1cm, single arrow, single arrow head extend=.4cm, rounded corners=0.05cm]
\tikzstyle{plate1}=[rectangle, fill=plate1, rounded corners=0.5cm]
\tikzstyle{plate2}=[rectangle, fill=plate2, rounded corners=0.5cm]
\tikzstyle{plate1lab}=[text=plate1lab, label distance=-7mm]
\tikzstyle{plate2lab}=[text=plate2lab, label distance=-7mm]

\pgfplotsset{
       myAxis/.style={
        ybar, axis on top,
        height=4cm, width=5cm,
        bar width=0.09cm,
        enlarge y limits={value=.1,upper},
        ymin=0, ymax=1,
        axis x line*=bottom,
        y axis line style={opacity=0},
        yticklabels={,,},
        tickwidth=0pt,
        enlarge x limits=true,
        xtick=data
    }
}

\tikzstyle{layer}=[top color=layerT, bottom color=layerB, general shadow=myShadow, rounded corners=0.1cm, minimum size=0.8cm]
\tikzstyle{parameter}=[top color=parameterT, bottom color=parameterB, general shadow=myShadow, rounded corners=0.1cm]
\tikzstyle{vmissing}=[draw=none, scale=2,text height=0.333cm,execute at begin node=\color{black}$\vdots$]
\tikzstyle{hmissing}=[draw=none, scale=2,text width=0.41cm,execute at begin node=\color{black}$\cdots$]
\tikzstyle{arrow}=[->, >=stealth]
\tikzstyle{arrowR}=[<-, >=stealth]

\tikzstyle{bias}=[circle,draw]
\tikzstyle{operation}=[circle,draw]
\tikzstyle{activation}=[draw]
\tikzstyle{delay}=[draw,minimum size=0.5cm,fill=black]
\tikzstyle{line}=[]
\tikzstyle{vectorLine}=[line width=0.6mm]
\tikzstyle{vectorArrow}=[->, >=stealth, line width=0.6mm]
\tikzstyle{border}=[draw]

\pgfplotstableset{
    color cells/.style={
        col sep=comma,
        string type,
        postproc cell content/.code={%
                \pgfkeysalso{@cell content=\rule{0cm}{2.4ex}\cellcolor{rgb:white,10;red,\xintRound {4}{\xinttheexpr 100*##1\relax};blue,\xintRound {4}{\xinttheexpr 10-(10*##1)\relax}}##1}%
                },
        columns/x/.style={
            column name={},
            postproc cell content/.code={}
        },
        columns/35/.style={column name=$\mathcal{P}_{t_k}^{(35)}$\vspace{1mm}},
        columns/36/.style={column name=$\mathcal{P}_{t_k}^{(36)}$},
        columns/37/.style={column name=$\mathcal{P}_{t_k}^{(37)}$},
        columns/38/.style={column name=$\mathcal{P}_{t_k}^{(38)}$},
        columns/39/.style={column name=$\mathcal{P}_{t_k}^{(39)}$},
        columns/40/.style={column name=$\mathcal{P}_{t_k}^{(40)}$},
    }
}

\newcommand{\tc}[1]{\FPeval{\result}{clip((#1-49.5)/(97-49.5)*100)}\textcolor{red!\result!blue}{#1}}

\begin{document}

\title{Machine learning classification of non-Markovian noise disturbing quantum dynamics}

\author{Stefano Martina}
\email{stefano.martina@unifi.it}
\affiliation{Dept.\,of Physics and Astronomy, University of Florence, via G. Sansone 1, 50019 Sesto Fiorentino, Italy.}
\affiliation{European Laboratory for Non-Linear Spectroscopy (LENS), University of Florence, via N. Carrara 1, 50019 Sesto Fiorentino, Italy.}

\author{Stefano Gherardini}
\email{gherardini@lens.unifi.it}
\affiliation{CNR-INO, Area Science Park, Basovizza, I-34149 Trieste, Italy}
\affiliation{Dept.\,of Physics and Astronomy, University of Florence, via G. Sansone 1, 50019 Sesto Fiorentino, Italy.}
\affiliation{European Laboratory for Non-Linear Spectroscopy (LENS), University of Florence, via N. Carrara 1, 50019 Sesto Fiorentino, Italy.}

\author{Filippo Caruso}
\email{filippo.caruso@unifi.it}
\affiliation{Dept.\,of Physics and Astronomy, University of Florence, via G. Sansone 1, 50019 Sesto Fiorentino, Italy.}
\affiliation{European Laboratory for Non-Linear Spectroscopy (LENS), University of Florence, via N. Carrara 1, 50019 Sesto Fiorentino, Italy.}


\begin{abstract}
In this paper machine learning and artificial neural network models are proposed for the classification of external noise sources affecting a given quantum dynamics. For this purpose, we train and then validate \emph{support vector machine}, \emph{multi-layer perceptron} and \emph{recurrent neural network} models with different complexity and accuracy, to solve supervised binary classification problems. As a result, we demonstrate the high efficacy of such tools in classifying noisy quantum dynamics using simulated data sets from different realizations of the quantum system dynamics. In addition, we show that for a successful classification one just needs to measure, in a sequence of discrete time instants, the probabilities that the analysed quantum system is in one of the allowed positions or energy configurations. Albeit the training of machine learning models is here performed on synthetic data, our approach is expected to find application in experimental schemes, as e.g.\,for the noise benchmarking of noisy intermediate-scale quantum devices. 
\end{abstract}

\maketitle

\acrodef{ml}[ML]{Machine Learning}
\acrodef{nlp}[NLP]{Natural Language Processing}
\acrodef{svm}[SVM]{Support Vector Machine}
\acrodef{rbf}[RBF]{Radial Basis Function}
\acrodef{svc}[SVC]{Support Vector Classifier}
\acrodef{mmc}[MMC]{Maximal Margin Classifier}
\acrodef{ann}[ANN]{Artificial Neural Network}
\acrodef{gru}[GRU]{Gated Recurrent Unit}
\acrodef{lstm}[LSTM]{Long Short Term Memory}
\acrodef{rnn}[RNN]{Recurrent Neural Network}
\acrodef{cnn}[CNN]{Convolutional Neural Networks}
\acrodef{mlp}[MLP]{Multi-Layer Perceptron}
\acrodef{sgd}[SGD]{Stochastic Gradient Descent}
\acrodef{adam}[ADAM]{Adaptive Moment Estimation}
\acrodef{relu}[ReLU]{Rectified Linear Unit}

\section{Introduction}\label{sec:intro}

Noise sensing aims at discriminating, and possibly reconstructing, noise profiles that affect static parameters and dynamical variables governing the evolution of classical and quantum systems \cite{ColeNanotech2009,BylanderNatPhys2011,DegenRMP2017,SzankowskiJPCM2017}. In the quantum regime, which constitute the main object of our discussion, noise partially destroys the coherent evolution of the investigated open quantum system, interacting with an external environment or simpler with other systems \cite{BreuerBook2002,CarusoRMP2014}. In such scenario, noise can be generally modelled as a stochastic process, distributed according to an unknown probability distribution \cite{MuellerSciRep2016,Gherardini_Thesis}. As concrete examples, one may consider the following cases that have recently studied experimentally: (i) Resonant microwave fields with random amplitude and phase for the driving of atomic transitions \cite{DoNJP2019}; (ii) solid-state spin qubits in negatively charged nitrogen-vacancy (NV) centers that are naturally affected by a carbon nuclear spin environment \cite{HernandezPRB2018}; (iii) single photons undergoing random polarisation fluctuations \cite{KofmanPRA2001,VirziArXiv2021}. In all these experiments, noise stochastic fields sampled from an unknown probability distribution have to be included in the microscopic derivation of the system dynamics under investigation, in order to properly carry out noise sensing and discrimination.

Several techniques, at both the theoretical and experimental side, have been developed for the inference of the unknown noise distribution and to detect, if present, non-zero time-correlations among adjacent samples (over time) of the noise process \cite{Paz-SilvaPRL2014,BallPRA2016,NorrisPRL2016,FreyNatComm2017,MuellerSciRep2018,HernandezPRB2018,SungNatComm2019,DoNJP2019,KrzywdaNJP2019,niu2019learning,MuellerPLA2020,YoussryArXiv}. However, most of them suffer of the need to control the quantum system, by generating multiple control sequences (e.g., dynamical decoupling ones \cite{AlvarezPRL2011,YugePRL2011,PoggialiPRX2018}), each of them being sensitive to a different component of the noise spectrum \cite{CywinskiPRA2014,DallaPozzaIJQI2019}. In this regard, in Ref.\,\cite{HarperNatPhys2020} a diagnostic protocol for the detection of correlations among arbitrary sets of qubits have been tested on a 14-qubit superconducting quantum architecture, by discovering the persistent presence of long-range two-qubit correlations. Moreover, \ac{ml}-models have been also adopted to study non-Markovian open quantum dynamics \cite{luchnikov2020machine,fanchini2020estimating,wise2021using}. In particular, in \cite{luchnikov2020machine} a method is developed to learn the effective Markovian embedding of a non-Markovian process. The embedding is learned by maximising the likelihood function built over successively observed measurements of the quantum dynamics. The assumption in \cite{luchnikov2020machine} is that the underlying time-evolution of the system is non-Markovian and the focus of the work is the training of the Markovian embedding. Thus, it is not directly addressed the issue of discriminating the presence of noise sources affecting the dynamics, nor if the noise samples over time are time-correlated. Instead, in \cite{fanchini2020estimating} a \ac{svm} model is trained to predict the degree of non-Markovianity in open quantum systems. An open quantum system approach is thus employed, but without providing emphasis on quantum dynamics perturbed by a stochastic process of noise, nor on the use of more complex \ac{ml}-models as neural networks and \ac{rnn}. In Ref.\,\cite{wise2021using} a deep neural network approach is adopted to perform (at the theoretical level) noise regression of qubits immersed in their environment that entails different stationary, Gaussian noise spectra. In \cite{wise2021using}, deep neural networks are trained with time-dependent coherence decay (echo) curves used as input data. 

In this paper, differently to all the aforementioned references, we exploit \ac{ml} techniques \cite{ShaiUML2014,HastieESL2009} to efficiently carry out high accuracy classification of noise affecting quantum dynamics. The proposed methods are designed to distinguish between Independent and Identically Distributed (i.i.d.) noise sequences and noise samples originated by a non-trivial memory kernel, thus characterised by specific time-correlation parameters. It is worth reminding that, in the latter case, the dynamics of the stochastic quantum system (stochastic due to the presence of fluctuating parameters, e.g.\,in the Hamiltonian of the analyzed system as in \cite{DoNJP2019}) turns out of being non-Markovian \cite{RivasRPP2014,BreuerRMP2016}, in the sense that samples of its state in different time instants are correlated \cite{LupkePRXQuantum2020}. This entails that the propagation of the system in subsequent time intervals is highly influenced by its previous states, even occurring in the early stages of the dynamics \cite{PollockPRA2018,MilzPRXQuantum2021,GherardiniQST2022}. This effect corresponds to a two-fold exchange of information between the system and the external sources, which has thus applications for quantum sensing \cite{GiarmatziQuantum2021,FigueroaPRX2021}.

\begin{figure}[!ht]
  \centering
  \includegraphics[width=0.8\columnwidth]{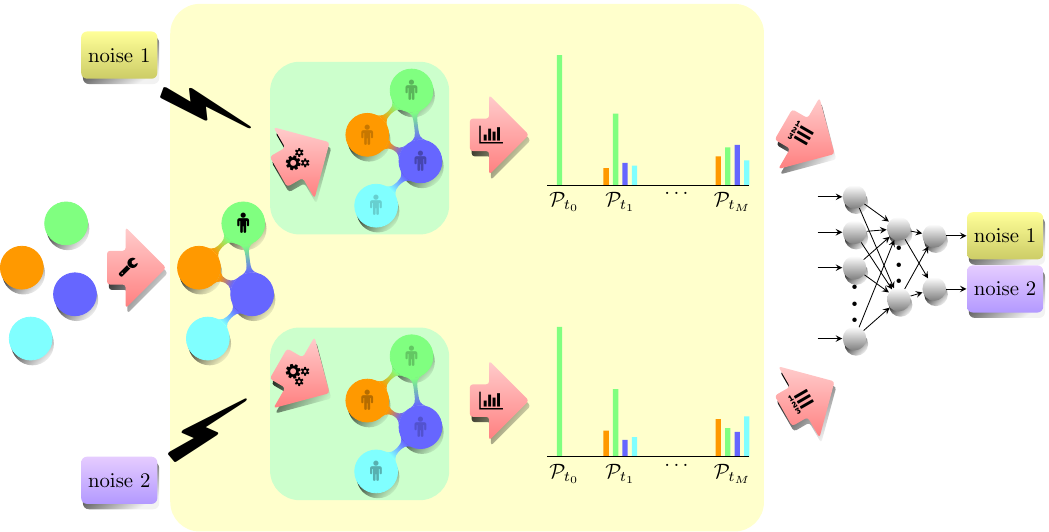}
  \caption{Pictorial representation of the proposed machine learning procedure for noise classification. For a fixed set of nodes $\mathcal{N}$ (coloured circles) we take into account the stochastic evolution of a quantum particle in a network affected by different types of noise sources. Such noisy quantum dynamics are evaluated at $M$ consecutive steps (small green plate). To make the synthetic generated data closer to a possible real setting, the topology $\mathcal{E}$ (edges linking the coloured circles) of the network and the initial state distribution $\mathcal{P}_{t_0}$ (black pawn on the green right circle) are chosen randomly for a predefined number of different configurations (yellow background). After the dynamics, all the distributions $\mathcal{P}_{t_k}$, in correspondence of the $M+1$ time instants $t_k$ with $k=0,\ldots,M$, are collected and recorded along with the noise type label. Then, a data set of $N$ different realisations is used to train a \ac{ml}-model (a neural network in the figure) for the classification of noise sources.}
  \label{fig:schema}
\end{figure}

To present our novel approach and demonstrate its efficacy in discriminating Markovian and non-Markovian noise sources, we focus on the dynamics of a quantum particle randomly moving on a graph $\mathcal{G}$ \cite{KempeCP2003,VenegasQIP2012,DallaPozzaPRR2020}, as generated by a stochastic Schr\"odinger equation. Depending on the way the particle is affected by the external noise, the noise-dependent component of its movements within the graph may be time-correlated. In this general context, we are going to propose \ac{ml}-based solutions for the classification of characteristic noise features. Specifically, by training a properly-designed \acl{ml} model via the probabilities that the particle is in each node of the graph $\mathcal{G}$ at discrete time instants 
%
%
(thus, no coherence decay curves need to be measured as in \cite{wise2021using}), we will show that it is possible to discriminate accurately between different noise sources and identify the possible presence of time-correlations from observation of the quantum system dynamics.   

To perform noise classification, \acp{svm}, \acp{mlp} and \acp{rnn} \cite{BishopPRML2006,GoodfellowDL2016,schmidhuber2015deep,goldberg2017neural} are successfully trained on six data sets (each of them composed of $20\,000$ realisations) that have been properly generated to carry out binary classification of noisy quantum dynamics. Once trained, the proposed \ac{ml}-models are able to reach a classification accuracy (defined by the number of correctly classified realisations over their total number) up to $97\%$. A pictorial representation of the proposed ML procedure is depicted in Fig.\,\ref{fig:schema}.

As other existing sensing techniques, the training of our \ac{ml}-models can be performed preliminary on synthetic data. Specifically, synthetic data are generated by solving a stochastic Schr\"odinger equation -- modeling the noisy quantum dynamics we are analyzing -- that exhibits at least one random parameter to be randomly sampled. As a result, we have observed that both i.i.d.\,and correlated noise sources can be accurately discriminated by means of one single \ac{ml} architecture. Moreover, our \ac{ml}-based approach allows for non-Markovian noise classification by processing only measurements of the diagonal elements (even called ``populations'') of the density operator $\rho_t$ associated with the quantum system under investigation. Thus, no measurements of the off-diagonal elements of $\rho_t$, stemming from quantum coherence terms in a given basis of interest, might be required. For example, for the quantum particle case, this means that we just need to record, in discrete time instants, the probabilities (denoted as ``occupation probabilities'') that the particle is in the positions (even part of them) identified by the nodes of the graph $\mathcal{G}$. These advantages can find application in experimental setups affected by stochastic noise sources as the ones in \cite{HernandezPRB2018,DoNJP2019,VirziArXiv2021}, and even in the available or coming quantum devices where a noise certification could be crucial before performing any task \cite{BallPRA2016,MorrisArxiv2019} (see also the subsection \ref{sec:proposal} below).

\section{Stochastic quantum dynamics}\label{sec:dynamic}

Let us introduce the general physical framework to which our \ac{ml} methods will be applied. For this purpose, we consider a quantum particle that randomly moves on a complex graph $\mathcal{G}$ by following the quantum mechanics postulates. The complex graph is described by the pair $(\mathcal{N},\mathcal{E})$, where $\mathcal{N}$ is the set of nodes or vertices while $\mathcal{E}$ is the set of links, denoted as $s \leftrightarrow \ell$, coupling pairs of nodes, with $s,\ell=1,\ldots,d$ and $d$ being the total number of nodes. Each node is associated with a different particle position, while the links correspond to the possibility that the particles jumps from a node to another. In particular, the links in $\mathcal{E}$ can be summarised in the adjacency matrix $A_t$ (time-dependent operator in the more general case), whose elements are given by
\begin{equation}
    A^{(s,\ell)}_t \equiv
    \begin{cases}
    g_t\,\,\,\,\text{if}\,\,\,\,
    s \leftrightarrow \ell \in \mathcal{E}  \\
    0\,\,\,\,\text{if}\,\,\,\,s \leftrightarrow \ell \not\in \mathcal{E}.  \\
    \end{cases}
\end{equation}
In this way, we are implicitly assuming that all the links are equally coupled with the same weight equal to $g_t$ that is taken as a time-dependent parameter. 

Here, the coupling $g_t$ is modelled as a stochastic process defined by the collection of random variables $\boldsymbol{g} \equiv \left(g_{t_0},\ldots,g_{t_{M-1}}\right)^T$, with $(\cdot)^{T}$ being the transposition operation, in correspondence of the discrete time instants $t_k$, $k=0,\ldots,M-1$. At each $t_k$, $g_t$ is sampled from a specific probability distribution ${\rm Prob}(g)$ and is assumed to remain constant at the extracted value for the entire time interval $[t_k,t_{k+1}]$. For simplicity, also the value $\Delta\equiv t_{k+1}-t_k$ is taken constant for any $k=0,\dots,M-1$, and the stochastic process $g_t$ is considered to take $D$ different values $g^{(1)},\ldots,g^{(D)}$ with probabilities $p_{g^{(1)}},\ldots,p_{g^{(D)}}$. In this way,
\begin{equation}
    {\rm Prob}(g) = \sum_{j=1}^{D}p_{g^{(j)}}\delta(g-g^{(j)})
\end{equation}
is provided by a discrete probability distribution with $D$ values, with $\delta(\cdot)$ denoting the Kronecker delta. 

If $\boldsymbol{g}$ is provided by a collection of i.i.d.\,random variables sampled from the probability distribution ${\rm Prob}(g)$, then the noise sequence that affects the link strength $g$ is uncorrelated over time, and it is denoted as \emph{Markovian}. Conversely, in case the occurrence of the random value $g^{(j)}$, $j=1,\ldots,D$, at the discrete time instants $t_{k}$, $k=0,\ldots,M-1$, depends on the sampling of $g$ at previous time instants, the noise sequence is time-correlated and the noise is denoted as \emph{non-Markovian} or as a \emph{coloured noise process}. In this regard, notice that the value of the parameters, which define the correlation among different samples of noise in single time-sequences, uniquely set the colour of the noise. Also observe that, known the multi-times distribution ${\rm Prob}(\boldsymbol{g})$ defined over the discrete time instants $t_{k}$, one can compute the noise auto-correlation function, whose Fourier transform is by definition the power spectral density of the noise process. In other terms, there is a one-to-one mapping between the representations of the noise in the time and frequency domains respectively. This entails that noise sensing can be performed in one of the two domain at best convenience. Moreover, this also motivates the generality of the stochastic quantum model we are here introducing that, indeed, can be applied to all those problems concerning the transport of single particles within a network \cite{DErricoNatComm2013,VicianiSciRep2016,HarrisNatPhot2017}, but also to quantum system dynamics influenced by the external environment as those in Refs.\,\cite{BylanderNatPhys2011,SzankowskiJPCM2017,SungNatComm2019}. 

In our model we adopt as correlation model the well-known formalism of time-homogeneous \emph{discrete Markov chains} \cite{StocProc_book2013}. The latter can be graphically interpreted as state-machines that assign the conditional probability of ``hopping'' from each possible value of $g$ to an adjacent one at consecutive time instants. Each conditional probability is defined, at any time $t$, by a \emph{transition matrix} $T$ that is a left or right stochastic operator. Let us remind that discrete Markov chains differ by a parameter $m$ named the \emph{order} of the chain. In a Markov chain of order $m$, future realisations of the sampled random variable (e.g., our $g_t$) depend on the past $m$ realisations in previous time instants. Here, we will consider $(m=1)$-order discrete Markov chains, namely correlated noise sequence characterised by a single (1-step) transition matrix $T$ that we aim to discriminate by means of properly-developed \ac{ml} techniques. This choice is simply dictated by our desire to effectively illustrate the obtained results, and not by intrinsic limitations of the methods we are going to propose. As an example, let us assume $m=1$ and $D=2$. In this specific case, by taking the conditional probabilities $p(g_{t_k}|g_{t_{k-1}})$ with $g_{t_k}$ equal to $g^{(1)}$ or $g^{(2)}$ for any $k$, it holds that $p(g_{t_k}|g_{t_{k-1}})$ is equal to one of the elements within the following transition matrix: 

\begin{small}
\begin{equation}
    T = \begin{pmatrix} p(g_{t_k}=g^{(1)}|g_{t_{k-1}}=g^{(1)}) & p(g_{t_k}=g^{(1)}|g_{t_{k-1}}=g^{(2)}) \\
    p(g_{t_k}=g^{(2)}|g_{t_{k-1}}=g^{(1)}) & p(g_{t_k}=g^{(2)}|g_{t_{k-1}}=g^{(2)})
    \end{pmatrix}.
\end{equation}
\end{small}

\noindent
Thus, the stochastic realisations of $g$ in different time instants are not correlated only if all the elements of $T$ are equal to $1/2$. In addition, we assume that all the nodes of the graph $\mathcal{G}$ have the same energy. Without loss of generality, one is allowed to set such energy to zero, with the result that the Hamiltonian $H_t$ of the quantum particle is identically equal to the adjacency matrix $A_t$, i.e., $H_t = A_t$ for any time instant $t$. Moreover, we consider that the state of the particle, moving on a graph with $d$ nodes, is provided by the density operator $\rho_t$ that, by definition, is an Hermitian, positive semi-definite, idempotent operator matrix with trace $1$. By using the vectorisation operation ${\rm vec}[\cdot]$, we convert $\rho_t$ into the column vector
\begin{eqnarray*}
&\boldsymbol{\lambda}_t \equiv {\rm vec}[\rho_t] & \\
& = (\rho_t^{(11)},\ldots,\rho_t^{(d1)},\rho_t^{(12)},\ldots,\rho_t^{(d2)},\ldots,\rho_t^{(dd)})\in\mathbb{C}^{d^2}& 
\end{eqnarray*}
where $\rho_t^{(s,\ell)}$ denotes the $(s\,\ell)$-element of $\rho_t$. The state $\boldsymbol{\lambda}_t$ is a vector of $d^2$ elements belonging to the space of complex numbers. Since a quantum particle can live in a superposition of positions, whereby also quantum coherence plays an active role, $d$ elements of $\boldsymbol{\lambda}_t$ corresponds to the probabilities of measuring the particle in each of the allowed positions, while the other elements are \emph{quantum coherence} terms that identify interference patterns between the nodes of the graph. Thanks to the vectorisation of $\rho_t$, the ordinary differential equation, governing the dynamics of the particle, is recast in a linear differential equation for $\boldsymbol{\lambda}_t$, i.e.,
\begin{eqnarray*}
    &&\frac{\partial}{\partial t}\boldsymbol{\lambda}_t
    =\mathcal{L}_t\,\boldsymbol{\lambda}_t \Longleftrightarrow \boldsymbol{\lambda}_t = e^{\mathcal{L}_t}\boldsymbol{\lambda}_0 \\
    &&\mathcal{L}_t \equiv -\frac{i}{\hbar}\left(\mathbb{I}_{d} \otimes A_t - A_{t}^T \otimes \mathbb{I}_d\right) 
\end{eqnarray*}
with $\otimes$ Kronecker product and $\hbar$ reduced Planck constant. By construction, $\mathcal{L}_t$ is a skew-Hermitian operator for any time instant $t$, i.e., $\mathcal{L}^{\dagger}_t$+$\mathcal{L}_t=0$ $\forall t$. 

\section{Problem formulation}\label{sec:problem_formulation}

Our aim is to identify the presence of noise sources acting on the coupling $g_t$ of the adjacency matrix $A_t$, and then discriminate among different noise probability distributions ${\rm Prob}(g)$ and correlation parameters in the samples of the time-sequences $\boldsymbol{g}$. Moreover, we also aim to evaluate if such tasks can be carried out by only measuring the population terms of the particle at the discrete time instants $t_k$, even by taking into account few runs of the quantum system dynamics. 

The population values are collected in the vectors $\mathcal{P}_{t_k}\in\mathbb{R}^{d}$ that have as many elements as the nodes of the graph. After each stochastic evolution of the quantum particle, $\mathcal{P}_{t_k}$ takes different values depending on the specific realisation of $\boldsymbol{g}$. 

At the experimental level, the population distributions $\mathcal{P}_{t_k}$ can be obtained in multiple runs, by stopping the stochastic evolution of the system at each time $t_k$ (with $k=1,\ldots,M$), then collecting the measurement records and restarting from the beginning the experimental routine. This means that one does not need to experimentally implement sequential measurements routines, requiring to take into account also the quantum measurement back-action on the state of the system. 
The measurement outcomes can be just recorded at the end of the quantum system evolution; however, this can be realized at the price of performing multiple runs of the stochastic quantum dynamics under scrutiny.

\subsection{Data set generation}\label{sec:dataset}

For the generation of the data used to train the \ac{ml}-models, we consider two variants of three different classification problems. Each sample of the data sets is created by first generating a random set of links $\mathcal{E}$ (random topology) for the graph $\mathcal{G}$, and then initialising the particle in a randomly chosen node of the graph. We set $M=15$ as the number of evaluations (measurements) of the quantum particle dynamics, and $d=40$ as the number of nodes of the graph $\mathcal{G}$. This means that $\mathcal{P}_{t_0}$ is a Kronecker delta centered in one of the 40 nodes, and the stochastic quantum dynamics is evolved for 15 steps for each simulated noise source of the generated data set. Here, it is worth noting that the choice of $M=15$ is dictated by the fact that in recent experiments as for instance in Refs.\,\cite{PiacentiniNatPhys2017,DoNJP2019,HernandezPRR2020}, the number of intermediate quantum measurements does not exceed $10$, and thus $M=15$ is sufficiently large to represent actual physical setups. Instead, regarding taking $d=40$, such a value is just able to generate a complex landscape for the particle dynamics and small enough to be numerically manageable. 
The total considered dynamical time $t_{M}$ is taken equal to $t_M=1$ or $t_M=0.1$ in dimensionless units, each of them corresponding to a specific variant. Notice, indeed, that the values of $t_{M}$ are expressed consistently with the energy scale of the couplings $g_t$, whose random values $g^{(j)}$ belong to the set $\{1,2,3,4,5\}$ in the data set generation, such that $\hbar$ can be reliably set to $1$ as usual. All the probability distributions $\mathcal{P}_{t_k}$ for $k=0,\dots,15$ are stored together with the attached label that indicates the associated type of noise.

For each of the two variants of our classification problems, we generate three different balanced data sets of $20\,000$ samples. The first data set, which we call \iid{}, is suitable for a supervised binary classification task that discriminates between two different i.i.d.\,noisy quantum dynamics, where the noise sources have the same support but different probability distribution ${\rm Prob}(g)$. 

The second data set, named as \nm{}, concerns the classification of two different coloured noisy quantum dynamics with the noise sources again having the same support but different ${\rm Prob}(g)$ (the same ones as in the data set \iid{}) and a transition matrix $T$. 

Finally, the third data set, called \vs{}, is created for the classification between stochastic quantum dynamics affected respectively by an i.i.d.\,and a coloured noise with same support and ${\rm Prob}(g)$. 

Note that choosing graphs with random links allows to increase the statistical variability of the input data, with the result that the \ac{ml} algorithms learn to classify noise sources independently of the graph topology. The aim, indeed, is to prevent that the ML-models rely only on features specific to a small class of topologies. Moreover, taking random initial distributions $\mathcal{P}_{t_0}$ allows to increase the \emph{robustness} of the \ac{ml} methods, making them less likely to overfit on the synthetic data set.

\begin{table}[t!]
\caption{Example of a part of $\mathcal{P}_{t_k}$ for all the discrete time instants $t_k$ for a noisy quantum dynamics affected by i.i.d.\,noise sources and $t_{15}=0.1$ (in dimensionless units). In the Table, $\mathcal{P}_{t_k}^{(s)}$ denotes the $s$-th element of the vector $\mathcal{P}_{t_k}$ for any $t_k$, $k=0,\ldots,15$.\\}
\label{tab:example1}
\centering
\pgfplotstabletypeset[color cells]{
x,35,36,37,38,39,40
$t_0$,0.00,0.00,0.00,0.00,1.00,0.00
$t_1$,0.00,0.00,0.00,0.00,0.99,0.00
$t_2$,0.00,0.00,0.00,0.00,0.93,0.00
$t_3$,0.00,0.00,0.01,0.01,0.85,0.01
$t_4$,0.00,0.01,0.01,0.01,0.78,0.01
$t_5$,0.01,0.01,0.01,0.01,0.69,0.01
$t_6$,0.01,0.02,0.01,0.01,0.63,0.00
$t_7$,0.01,0.02,0.01,0.01,0.57,0.00
$t_8$,0.01,0.02,0.01,0.01,0.52,0.00
$t_{9}$,0.02,0.02,0.01,0.01,0.45,0.00
$t_{10}$,0.02,0.02,0.02,0.02,0.37,0.01
$t_{11}$,0.01,0.02,0.02,0.03,0.29,0.01
$t_{12}$,0.01,0.01,0.03,0.04,0.20,0.02
$t_{13}$,0.01,0.01,0.04,0.04,0.14,0.01
$t_{14}$,0.01,0.02,0.04,0.05,0.08,0.01
$t_{15}$,0.01,0.02,0.04,0.05,0.06,0.01
}
\end{table}

\begin{table}[t!]
\caption{Example of a part of $\mathcal{P}_{t_k}$ for all the discrete time instants $t_k$ for a noisy quantum dynamics affected by i.i.d.\,noise sources and $t_{15}=1$ (in dimensionless units). Again, $\mathcal{P}_{t_k}^{(s)}$ denotes the $s$-th element of the vector $\mathcal{P}_{t_k}$ for any $t_k$, $k=0,\ldots,15$. The topology and the initial state, for this example, are the same of those in \Cref{tab:example1}.\\}
\label{tab:example2}
\centering
\pgfplotstabletypeset[color cells]{
x,35,36,37,38,39,40
$t_{0}$,0.00,0.00,0.00,0.00,1.00,0.00
$t_{1}$,0.02,0.02,0.01,0.01,0.45,0.00
$t_{2}$,0.01,0.01,0.02,0.03,0.05,0.02
$t_{3}$,0.00,0.00,0.00,0.00,0.13,0.02
$t_{4}$,0.02,0.01,0.01,0.01,0.12,0.02
$t_{5}$,0.01,0.01,0.03,0.03,0.06,0.01
$t_{6}$,0.01,0.01,0.01,0.01,0.01,0.00
$t_{7}$,0.04,0.03,0.01,0.06,0.11,0.00
$t_{8}$,0.04,0.00,0.03,0.11,0.11,0.03
$t_{9}$,0.03,0.00,0.03,0.01,0.01,0.10
$t_{10}$,0.05,0.01,0.01,0.04,0.08,0.04
$t_{11}$,0.01,0.03,0.02,0.00,0.08,0.02
$t_{12}$,0.00,0.05,0.02,0.04,0.00,0.06
$t_{13}$,0.01,0.03,0.00,0.02,0.05,0.07
$t_{14}$,0.00,0.00,0.00,0.01,0.12,0.00
$t_{15}$,0.00,0.00,0.01,0.04,0.10,0.01
}
\end{table}

As it will be explained in the following, some \ac{ml}-models that we are going to introduce will use as input only the last distribution $\mathcal{P}_{t_{15}}$, while other \ac{ml}-models will take all the $\mathcal{P}_{t_k}$ for any $t_k$. Moreover, each data set is balanced split in a \emph{training set} of $12\,000$ samples, a \emph{validation set} of $4\,000$ samples, and a \emph{test set} of $4\,000$ samples.

In \Cref{tab:example1,tab:example2} we plot the occupation probabilities $\mathcal{P}_{t_k}$ (just for the \iid{} case for the sake of an easier presentation), being here interested in looking for the difference between choosing $t_{15}=0.1$ or $1$, which identify the two different variants of the generated data set. In this regard, it is worth noting that the duration $t_{15}=0.1$ (in dimensionless units) of the quantum system dynamics, as in the example in \Cref{tab:example1}, is the minimal one to observe the diffusion of the system's population outside the node on which has been initialised. However, as it will be verified by our experiments and explained later, with this choice one has that, by taking $t_{15}=0.1$, the classification problem results quite straightforward. Indeed, just basic \ac{ml}-models that are only trained on $\mathcal{P}_{t_{15}}$ (thus, only on the final distribution $\mathcal{P}$) are able to correctly classify between two noisy quantum dynamics. Therefore, it was more interesting to increase the value of $t_{15}$ up to $t_{15}=1$ (in dimensionless units). As in the example of \Cref{tab:example2}, it leads to more complex data sets, and only deep learning models, designed to read all the $\mathcal{P}_{t_k}$, can classify the generated noisy quantum dynamics. 

As final remark, note that the current synthetic data set is build assuming perfect measurement statistics, as it was obtained from a large enough number of repetitions of the noisy quantum dynamics. Hence, to better adapt the synthetic data set to real data, one should simulate experimental case in which the measurement statistics are estimated from a finite number of dynamics realizations (i.e., measurement shots).

\subsection{Classification tasks}
We here present the binary supervised classification tasks that we are going to address, by taking $\mathcal{P}_{t_k}$ as input:
\begin{enumerate}[(i)]
\item 
Two different probability distributions ${\rm Prob}(g)$ -- specifically, $p_g^{(1)},\dots,p_g^{(5)}=(0.0124,\,0.04236,\,0.0820,\,0.2398,\,0.6234)$ and $=(0.1782,\,0.1865,\,0.2,\,0.2107,\,0.2245)$ -- both associated with i.i.d.\,noise sources. 
\item 
Two different ${\rm Prob}(g)$ (the same as (i)) and different values of the correlation parameters -- identified by transition matrices $T$ as explained in Sec.\,\ref{sec:dynamic} -- for coloured (thus, non-Markovian) noise processes. 
\item  
An i.i.d.\,and a coloured noise process with the same support $g$ and distribution $p_g^{(1)},\dots,p_g^{(5)}=(0.0124,\,0.04236,\,0.0820,\,0.2398,\,0.6234)$ that thus differ for the presence of non-zero correlation parameters.
\end{enumerate}
The values of both ${\rm Prob}(g)$ and the transition matrices $T$, used in our numerical simulations, are chosen randomly.

To solve the classification tasks above, we have employed in this paper both a more standardized \ac{ml}-model that is \acf{svm}\,\cite{HastieESL2009} and more recent \acp{ann}\,\cite{russakovsky2015imagenet,schmidhuber2015deep,krizhevsky2017imagenet,emmert2020introductory} models in the form of \acp{mlp} and \acp{rnn}.
For an exhaustive explanation of such a \ac{ml}-models refer to the Appendix.

\section{Results}\label{sec:results}

In our work, we consider two \ac{svm} models as baseline. The first one is denoted \svm{} and uses as input only the final probability distribution $\mathcal{P}_{t_{15}}$ (the prefix \emph{m-} stands for ``model'', to avoid confusion with the algorithm name; the suffix \emph{-single} means that it is based only on $\mathcal{P}_{t_{15}}$). Instead, the second one, which we call as \svmt{}, uses the set of all the $\mathcal{P}_{t_k}$ with $k=0,\dots,15$. For both of them, we try the following \emph{kernels} to increase the dimension of the feature-space that makes linearly separable the data-set: linear, polynomial with degree 2, 3 and 4, and \ac{rbf}. 

Then, we denote with \mlp{} a \ac{mlp} (also refer to \Cref{eq:mlp} in the Appendix), with $\ve{x}\equiv\mathcal{P}_{t_{15}}$ and $\ve{y}\equiv (0,1)$ or $(1,0)$ to identify the two noisy quantum dynamics that we aim to classify. Differently, \mlpt{} takes as input the set of all $\mathcal{P}_{t_k}$.

Moreover, \gru{} and \lstm{} are unidirectional \acp{rnn} that employ the final hidden representation (see \Cref{eq:aggregation,eq:rnnAll} in Appendix for more details). They are implemented by exploiting the \ac{gru} and \ac{lstm} methods, respectively.
The input to the models is $\ve{x}_{i+1}\equiv\mathcal{P}_{t_{i}}$, with $i=0,\dots,15$, while the output $\ve{y} \equiv (0,1)$ or $(1,0)$ as before. Besides, \bigru{} and \bilstm{} are the bidirectional versions of \gru{} and \lstm{}, while \bigrua{} and \bigrum{} are as \bigru{} but in addition, respectively, with an attention mechanism
and a max pooling (respectively, Eqs.\,(\ref{eq:aggAtt}) and (\ref{eq:aggMax}) in Appendix) as forms of aggregation of the \ac{rnn} hidden representations . Similarly, \bilstma{} and \bilstmm{} are the attentive and max pooling equivalents of \bilstm{}, respectively.

\begin{table}[!ht]
    \caption{Percent accuracy $\gamma$ (calculated on the test set) of the \ac{ml}-models trained in the tasks of binary classification of noisy quantum dynamics with: (i) Two different i.i.d.\,noise sources (\iid{}); (ii) two different coloured noise processes (\nm{}) leading to non-Markovian dynamics; and (iii) one i.i.d.\,vs one coloured noise sources (\vs{}). In this regard, let us recall that the coloured noise processes addressed in this paper are such that the probability distributions $\mathcal{P}_{t_k}$ depend both on ${\rm Prob}(g)$ and $1$-step transition matrix $T$.} In the first three columns of the table, the total duration of the dynamics is equal to $t_{15}=0.1$, while in the last three is $t_{15}=1$. The first two rows of the table report the results of the \ac{ml}-models that use as input only $\mathcal{P}_{t_{15}}$, while the models of the other rows take as input all the probability distributions $\mathcal{P}_{t_k}$ for $k=0,\dots,15$. The highest values of the accuracy have been underlined, and a color gradient (from blue to bright red) highlights the difference in their values.\\
    \label{tab:results}
    \centering
    \begin{tabular}{llccc|ccc}
        && \multicolumn{3}{c}{$t_{15}=0.1$}& \multicolumn{3}{c}{$t_{15}=1$} \\
        && \iid & \nm & \vs & \iid & \nm & \vs  \\
        \hline
        \multirow{2}{*}{\begin{minipage}[t]{1cm}
          $\mathcal{P}_{t_{15}}$
        \end{minipage}
        } & \svm & \underline{\tc{97.0}} & \underline{\tc{82.3}} & \tc{96.5} & \underline{\tc{50.3}} & \underline{\tc{51.2}} & \tc{49.5} \\
        &\mlp & \tc{96.9} & \tc{80.7} & \underline{\tc{96.6}} & \tc{49.5} & \tc{50.7} & \underline{\tc{50.2}} \\
        \hline
        \multirow{10}{*}{\begin{minipage}[t]{1cm}
          $\mathcal{P}_{t_0},$
          \endgraf
          $\dots,$
          \endgraf
          $\mathcal{P}_{t_{15}}$
        \end{minipage}
        } & \svmt & \tc{96.4} & \tc{80.1} & \tc{96.3} & \tc{73.6} & \tc{61.9} & \tc{75.0} \\
        &\mlpt & \tc{96.7} & \tc{80.7} & \tc{96.3} & \tc{74.0} & \tc{61.4} & \tc{70.7} \\
        &\gru & \tc{96.5} & \tc{91.5} & \underline{\tc{96.7}} & \tc{90.5} & \tc{73.3} & \tc{88.2} \\
        &\lstm & \tc{96.8} & \tc{90.4} & \tc{96.4} & \tc{88.6} & \tc{70.3} & \tc{86.3} \\
        &\bigru & \tc{96.6} & \tc{92.2} & \tc{96.6} & \tc{91.0} & \tc{74.6} & \underline{\tc{90.6}} \\
        &\bilstm & \tc{96.7} & \tc{89.7} & \tc{96.5} & \tc{90.8} & \tc{70.6} & \tc{87.2} \\
        &\bigrua & \underline{\tc{97.0}} & \tc{91.6} & \tc{96.1} & \tc{90.9} & \tc{73.4} & \tc{87.9} \\
        &\bilstma & \tc{96.9} & \tc{87.9} & \tc{96.3} & \tc{89.0} & \tc{71.6} & \tc{87.4} \\
        &\bigrum & \tc{96.6} & \underline{\tc{92.6}} & \tc{96.6} & \underline{\tc{91.8}} & \underline{\tc{76.1}} & \tc{90.4} \\
        &\bilstmm & \tc{96.6} & \tc{91.4} & \tc{96.3} & \tc{91.4} & \tc{74.9} & \tc{89.0} \\
        \hline
    \end{tabular}
\end{table}

In \Cref{tab:results}, for each model we report the best classification accuracy that is computed on the predictions performed over the test set. More formally, we define the prediction set
\begin{equation*}
    \Gamma \equiv \left\{(\ve{y}_1,\hat{\ve{y}}_1),\dots,(\ve{y}_n,\hat{\ve{y}}_n)\right\}
\end{equation*}
where $\ve{y}_1,\dots,\ve{y}_n$, taken from the data set, denote the true noise sources affecting the quantum system dynamics, and $\hat{\ve{y}}_1,\dots,\hat{\ve{y}}_n$ the corresponding predictions of the ML-model. Hence, the (percent) accuracy $\gamma$, function of $\Gamma$, is provided by
\begin{equation}
    \gamma(\Gamma)\equiv\frac{100}{n}\sum_{i=1}^n\indicator\left\{\argmax_{j=1,2} \hat{\ve{y}}_i^{(j)} = \argmax_{j=1,2} \ve{y}_i^{(j)}\right\},
\end{equation}
where 
\begin{equation}
    \indicator\{c\} \equiv \begin{cases}
    1, & \text{if}\,\,\,c\,\,\,\text{is true},\\
    0, & \mathrm{otherwise}.
    \end{cases}
\end{equation}
is the so-called \emph{indicator function}. The accuracy $\gamma$ defines the correctness of the model and can be used as a metric to identify which solution is better. In detail, for binary classification problems, as in our case, if $\gamma\simeq 50$ the classification is equivalent to perform a random guess, thus the model does not work. Instead, when $\gamma\simeq 100$ the model perfectly classifies all the elements of the test set, thus it is a nearly ideal classifier.

From the table, one can first observe that, by dealing with a total duration of the dynamics (in dimensionless units, by rescaling as the inverse of the couplings $g_t$) equal to $t_{15}=0.1$, we can reach the $97\%$ and $96.6\%$ of accuracy for the classification tasks \iid{} and \vs{} via an \ac{svm} using as input only the distribution $\mathcal{P}_{t_{15}}$. Instead, the task \nm{} is more difficult: $82.3\%$ of accuracy is achieved by \ac{svm}s applied just on $\mathcal{P}_{t_{15}}$. \ac{mlp} does not provide better results. In this case (\nm{} tasks), to obtain an accuracy over $90\%$, one can resort to \acp{rnn} taking as inputs all the $\mathcal{P}_{t_{k}}$ for $k=0,\dots,15$.

Conversely, for a longer dynamics, i.e., with $t_{15}=1$, we notice that using only $\mathcal{P}_{t_{15}}$ all the three classification tasks are not solved neither with \ac{svm} nor \ac{mlp}. Indeed, the accuracy $\gamma$ is always around $50\%$ and the models basically perform random guesses. The accuracy is increased by means of an \ac{svm} or an \ac{mlp} based on all $\mathcal{P}_{t_{k}}$, with $k=0,\dots,15$ as input. However, to get over $90\%$ of accuracy on the tasks \iid{} and \vs{}, we need to employ \acp{rnn}. The task \nm{} with $t_{15}=1$ is the most difficult among the analysed ones, and just $76.1\%$ of accuracy is obtained using \acp{rnn}. It is worth noticing that, for the tasks with $t_{15}=1$, we have empirically observed that the models adopting \ac{gru} perform better with respect to the ones that employ \ac{lstm}. Moreover, setting the bidirectionality in the \acp{rnn} allows slight improved accuracy, as well as the use of max pooling in aggregation. Instead, the attention mechanism does not seem to be beneficial for those tasks.

Among the proposed solutions, the more-performing is \bigrum{} that is realised by a bidirectional \ac{rnn} with \ac{gru} and max pooling aggregation. However, from our numerical simulations, we have observed that, independently on the employed \ac{ml}-model, the value of the total dynamical time as well as $M$ and $\Delta$ (see also paragraph \ref{subsec:scaling_class_accuracy}) emerge to be crucial for quantum noise classification. Specifically, by taking a quantum dynamics with a short enough duration, also \acp{svm} are able to classify quantum noise sources with very high accuracy. With short enough dynamics we mean short with respect to the time needed to the particle in escaping from the initial node of the graph, which in our case is around $t_{15}=0.1$. Instead, with $t_{15}$ around $1$ only \ac{rnn}s provide better results, and for $t_{15}\gg 1$ none of the proposed ML-techniques solves quantum noise classification problems (these results have not been reported in \Cref{tab:results} for the sake of better presentation). It is also worth stressing that, if the duration of the quantum dynamics is $t_{15}=0.1$, ML-models efficiently classify quantum noise sources by only processing the last measured distribution $\mathcal{P}_{t_{15}}$. These findings can be relevant for effective implementation (also at the experimental level), since the training and tuning of \ac{svm} is orders of magnitude faster with respect to \acp{ann} (e.g., around minutes vs hours or even days depending on the model and provided that a GPU is used). The reason to that has to be found in the more complex structure of the \acp{ann} than \acp{svm}.

\subsection{Scaling of the classification accuracy}\label{subsec:scaling_class_accuracy}

Let us now investigate the scaling of the classification accuracy $\gamma$, as a function of both the interval $\Delta$ between two consecutive transitions for $\ve{g}$ and the number $M$ of discrete time instants. Notice that $\Delta$ and $M$ are related to the total dynamical time $t_{M}$, since $t_{M}\equiv M\Delta$. 

A possible explanation of the differences observed between the three previously-analysed scenarios, i.e., $t_{15}=0.1$, $t_{15}=1$ and $t_{15}\gg 1$ (in dimensionless units), could be that the information on both the noise source and the initial quantum state 
is lost during the evolution of the system. For such aspect, not only the total dynamical time $t_{15}$ could play a role, but also the time interval $\Delta \equiv t_{1} - t_{0} \equiv \dots \equiv t_{M}-t_{M-1}$. In fact, it is reasonable to conjecture that a ML-model, able to correctly classify our noisy quantum dynamics with $t_{15}=1$ (thus $M=15$), can also work with $t_{M'}\gg 1$ for $M'>15$ and $\Delta'=\Delta$ where $\Delta'\equiv t_{1} - t_{0} \equiv \dots \equiv t_{M'}-t_{M'-1}$. In this way, the sequence $\mathcal{P}_{t_1},\dots,\mathcal{P}_{t_{15}}$ is contained in $\mathcal{P}_{t_1},\dots,\mathcal{P}_{t_{M'}}$. In other terms, we conjecture that the classification problem can be solved even for longer noisy quantum dynamics, but provided that $\Delta$ remains small.

To gain evidence on this conjecture, we have performed two additional experiments. Starting from the task \iid{} with $t_{15}=1$ and \bigrum{} as baseline (accuracy $91.8\%$), the same model (optimised in the same hyperparameters space) is trained on two new data sets. In both data sets, $t_{M}=2$ with $M$ equal to $15$ for the first data set and $30$ for the second one. Thus, in the former $\Delta'>\Delta$ with $\Delta$ time interval of the original data set, while in the latter $\Delta'=\Delta$. The first experiment ($\Delta'>\Delta$) provides a classification accuracy of $81.1\%$, contrarily to the results from the second experiment ($\Delta'=\Delta$), where a better accuracy of $96.3\%$ is achieved. We thus observe that, by taking $\Delta'=\Delta$ and the same ML-model, the classification problem can be solved with an higher accuracy, but at the price of a longer training time. Indeed, in this case, the length of each sample of the data set is twice the original one.

\begin{figure}[!ht]
  \centering
    \includegraphics[width=0.45\textwidth]{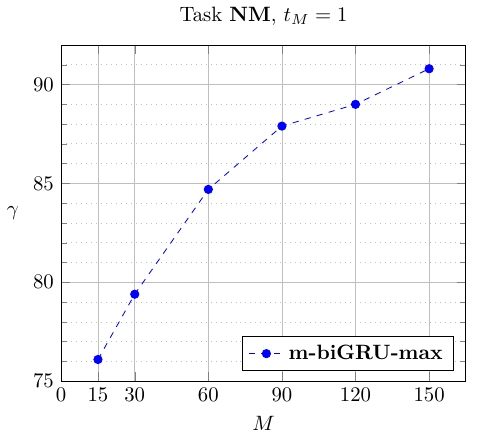}
  \caption{Percent classification accuracy $\gamma$ vs.\ $M$. It refers to the test set associated with the model \bigrum{} for the task \nm{}, where the value of the total evolution time is fixed to $t_M=1$. It is worth noting that the first point of the figure corresponds to the value in \Cref{tab:results} obtained for $t_{15}=1$.}
  \label{fig:plot}
\end{figure}
In another experiment, whose results are shown in \Cref{fig:plot}, we vary $M$ by keeping the total evolution time equal to $t_M=1$. Such tests use as baseline the model \bigrum{} applied on the most difficult task of \cref{tab:results}, i.e., \nm{} with $t_{15}=1$. As a result, the achieved classification accuracy is directly proportional to $M$ and, thus, inversely proportional to the value of the time interval $\Delta$. Indeed, by taking $t_M$ fixed and reducing $\Delta$, the classification accuracy of the same model can be enhanced. Specifically, it is possible to obtain more than $90\%$ of accuracy also for the task \nm{} with a total dynamical time equal to $1$, at the price of a longer training time as the length of the sequences increases.

\subsection{Quantum advantages}

Here, we address the following question: Could the proposed ML techniques be applied for the inference of noise sources affecting the dynamics of classical systems, e.g., Langevin equations\,\cite{DeGrootBook1984}? Probably yes, but we expect that their application to quantum systems, maybe surprisingly, can be more effective than on classical systems. Both classical (non-periodic) dissipative dynamics\,\cite{DeGrootBook1984} and stochastic quantum dynamics (stochastic due to the presence of an external environment, or noise sources as in our case) can asymptotically tend to a fixed-point, whereby the information on the initial state is lost. This means that the states of the system used for this noise classification tend to become indistinguishable as time increases. Classically, this can happen due to energy dissipation introduced by damping terms. Instead, quantum-mechanically, a dynamical fixed point can be reached due to decoherence that makes vanishing, at least on average, all quantum coherence terms\,\cite{BreuerBook2002}. Thus, once the transient of the evolution is elapsed, the evaluation of the final state of the system does not bring information neither on the initial state nor on the initial dynamics bringing the system to the asymptotic fixed-point. In our case, we have observed that, by using only $\mathcal{P}_{t_{15}}$ with long total dynamical time, the accuracy of all the classification tasks is always around $50\%$ both for \ac{svm} and \ac{mlp}. Consequently, if one aims to infer/reconstruct the value of parameters, signals or operators that influence the system dynamics by measuring its evolution, the most appropriate time window is during the transient. In this regard, a quantum dynamic, until it is nearly close of being unitary, is able to explore different configurations thanks to linearity and the quantum superposition principle. Conversely, classical dynamics, not being able to propagate superpositions of their trajectories, cannot provide per time unit the same amount of information on the quantity to be inferred.

In conclusion, the application of the proposed methods should be more accurate if applied to quantum systems than classical ones, but during the transient of its dynamical evolution when quantum effects are still predominant and the distance among the state and the fixed point is not negligible.

\subsection{Proposal for application to quantum computers}\label{sec:proposal}

Our techniques are expected to be adopted to witness the (non-)Markovianity of the noise sources in commercial quantum devices, as for example the Q-IBM\textsuperscript{\textregistered} \cite{MorrisArxiv2019} or Rigetti\textsuperscript{\textregistered}. In fact, such quantum devices, as other Noisy Intermediate-Scale Quantum prototypes \cite{preskill2018quantum}, are unavoidably affected by the external environment that entails random errors. Recently, in Refs.\,\cite{MartinaIBM2021,MartinaSoftwareImpacts}, it has been shown that it is possible to discriminate different quantum computers by looking at the (unknown) noise fingerprints that characterize each device. Thus, what the \ac{ml} techniques -- presented here -- could provide as an added value is to witness whether such noise fingerprints are time-correlated, and possibly how much the time-correlation is non-Markovian. For such experimental noise benchmarking, as in \cite{MartinaIBM2021}, it could be convenient to fix the connections among quantum gates (i.e., the underlying topology), and then consider more realizations of the implemented quantum dynamics affected by noise, so as to avoid the monitoring of the dynamics (cf.~Section~\ref{sec:problem_formulation}).  
Notice that to make a successful classification among i.i.d.\,and (non-)Markovian noise samples, one should be able to previously label them as Markovian or not Markovian, and more in general to understand the main features of the noise acting on the machines. However, such task is usually very hard to be carried out. Thus, as a more plausible strategy, we propose to compare the experimental data to the theoretical prediction at the level of multi-time measurement statistics, according to the following two steps:
\begin{itemize}
    \item[1)] Discriminate if and how much the measurement statistics (provided by the distributions $\mathcal{P}_{t_k}$ with $k=1,\ldots,M$, which have been measured on the real quantum devices on multiple runs) differ from the corresponding theoretical predictions. Such a difference, here on denoted as $\mathfrak{D}$, between theoretical and experimental data returns an effective prediction of the presence of noise on the machines. 
    \item[2)] Conditionally to step 1), evaluate with \ac{ml}-models the presence or the absence of functional relations $\mathcal{F}$ that link the difference distributions $\mathfrak{D}_{t_k}$ in correspondence of the time instants $t_k$. If two consecutive instances of $\mathfrak{D}$ at times $t_{k-1}$ and $t_k$ are no functionally related, then the noise is originated from a i.i.d.\,stochastic process. If, instead, there exist a functional $\mathcal{F}_{t_{k-1}:t_k} = f(\mathfrak{D}_{t_{k-1}},\mathfrak{D}_{t_{k}})$ that links together $\mathfrak{D}_{t_{k-1}}$ and $\mathfrak{D}_{t_k}$ in a non-trivial way, then the noise would come from a Markovian process. Finally, the noise process would be non-Markovian for functional relations $\mathcal{F}_{t_{k-n}:t_k}$ defined over multi times with $n>1$. \\
    The empirical characterization of $\mathcal{F}$, as well as the witness of non-Markovianity in the noise samples, can be obtained through the use of generative models. We can train three different models: (i) one model that generates $\mathfrak{D}_{t_k}$ by processing $\mathfrak{D}_{t_{k-n}},\dots,\mathfrak{D}_{t_{k-1}}$; (ii) another generative model that returns $\mathfrak{D}_{t_k}$ by taking $\mathfrak{D}_{t_{k-1}}$ as input; (iii) a model that directly generates $\mathfrak{D}_{t_k}$. If these three models have the same accuracy, then the noise process is likely i.i.d.. While, if the generative model (ii) has higher accuracy with respect to (iii) and the same accuracy than (i), then the noise process would be Markovian. Finally, if (i) has higher accuracy with respect to the models (ii) and (iii), then the noise process is non-Markovian.
\end{itemize}

To conclude, according to our proposal that will be tested in a forthcoming paper, time-correlations in the noisy samples of the distributions $\mathcal{P}_{t_k}$, with $k=1,\ldots,M$, can be determined by classifying functional relations $\mathcal{F}$ linking the difference distributions $\mathfrak{D}$, obtained by comparing the theoretical and measured values of $\mathcal{P}$ for a set of time instants. This is equivalent to discriminate coloured noise processes originated by different discrete Markov chain with non-zero transition matrices $T$. 
As a remark, it is still worth noting that also the experimental realization of the proposed procedure can be performed on multiple runs without the need to implement sequential measurements routines. Hence, each time a projective measurement is performed and the resulting measurement outcome recorded, the implemented (noisy) quantum circuit shall be executed from the beginning.

\section{Conclusions}

In this paper, we have addressed non-Markovian noise classification problems by means of deep learning techniques. In particular, the use of \ac{rnn} -- developed for sequence processing -- is motivated by the fact that we deal with time-ordered sequences of data. Even without resorting to external driving that may hinder detection tasks, we managed to classify with high accuracy stochastic quantum dynamics characterized by random parameters sampled from different probability distributions, associated with i.i.d.\,(Markovian) and coloured (non-Markovian) noise processes. For such a purpose, several \ac{ml}-models have been tested; in this regard, refer to \Cref{tab:results} for a summary of the results in term of the classification accuracy.

Among the proposed solutions, the more-performing is \bigrum{} that is realised by a bidirectional \ac{rnn} with \ac{gru} and max pooling aggregation. In fact, recurrent neural networks are particularly suitable to accomplish temporal machine-learning tasks thanks to their capability to generate internal temporal dynamics based on feedback connections. However, independently on the employed \ac{ml}-model, different accuracy values are achieved depending on the values of $M$, $\Delta$ and the total dynamical time. The way our \ac{ml} techniques rely on the parameters of the model has been addressed in the paragraph \ref{subsec:scaling_class_accuracy}.

Overall, all our numerical results have shown that it is easier to classify between two different noisy quantum dynamics both affected by i.i.d.\,noise sources or by i.i.d.\,and coloured noise processes than between two noisy quantum dynamics subjected to coloured noise. Again it confirms the relevant role played by time-correlations and how the latter highly influence the value of the classification accuracy. Furthermore, we also expect that the same ML-techniques that we have exploited in this work could be successfully applied to classify among coloured noise with $q$-step transition matrices $T_{t|t-q}$ with $q>1$.

\subsection{Outlooks}

As outlook, we plan to test the \ac{ml}-models employed in this paper on reconfigurable experimental platforms as the ones in Refs.\,\cite{nokkala2018reconfigurable,leedumrongwatthanakun2020programmable}, even affected by multiple noise sources. Moreover, we also aim to adapt our \ac{ml} methods (and especially \ac{ann}s) to reconstruct noise processes with time-correlation as key feature in the context of regression task instead of classification.
Indeed, our proposal is to provide accurate estimates of both the probability distribution ${\rm Prob}(g)$ and the transition matrix $T$, and the analysis would be extended for the
prediction of spatially-correlated noise sources. In this way, \ac{ml} approaches would represent a very promising, and possibly more accurate, alternatives to other noise-sensing techniques, e.g., those recently discussed in Refs.\,\cite{Paz-SilvaPRA2019,FreyPRApp2020}.

A well-known problem in \ac{ml} is the generalization to data shift. A model that is trained on a data set sampled from a specific data distribution will work correctly only with data sampled from the same distribution. In this paper, we used only synthetic data to evaluate the correctness of the training process and the \ac{ml} techniques. Thus, in order to validate this approach to real data, we should first collect them. This is out of the scope of the current work, but, as a remark, we can delineate three possible ways to build a real experimental data set. The first strategy is to acquire information a-priori on noise sources affecting the quantum system of interest in some experimental contexts by means of standard spectroscopy techniques, so that we can train the proposed \ac{ml}-models to discriminate between unseen classes of noise. In this way, the initial effort in building a training data set that also contains experimental data is counterbalanced by the possibility to predict noise features by means of faster classification tasks. The second strategy, which has been employed in \cite{MartinaIBM2021}, is to collect experimental data that comes from different noisy measurement statistics whose noise processes are not necessarily known. The \ac{ml}-models, then, are trained to classify (unknown) noise sources in distinct unseen sets of measurements. Finally, the third strategy, which is aimed to reduce the effort in building an informative experimental data set, is to train the \ac{ml}-models first on synthetic data and then to fine tune the training on a smaller further data set with only experimental data. In such a case, it is beneficial to adopt a synthetic data set that closely adapts to the real experimental setup. For instance, a simulated extra error can be added to the measurement statistics (in our paper provided by the distributions $\mathcal{P}_{t_k}$, $k=1,\ldots,M$) to take into account the finite number of measurement shots used for their computation.

\section*{Acknowledgments}

The authors were financially supported from by the Fondazione CR Firenze through the project QUANTUM-AI, the European Union’s Horizon 2020 research and innovation programme under FET-OPEN Grant Agreement No.\,828946 (PATHOS), and from University of Florence through the project Q-CODYCES.

\section*{Data and code availability}

The source codes for the generation of the data sets and the \ac{ml} experiments are available on \emph{GitHub} at the following address: \\
{\fontsize{8.7pt}{20pt}\selectfont\url{https://github.com/trianam/quantumNoiseClassification}}.

\section*{Appendix: Details on the employed ML models}

In this section, aiming at addressing also an audience not necessarily expert in \ac{ml}, we describe more in detail the \ac{ml}-models used in our tasks.

A generic binary data set in input to \ac{ml}-models is usually represented by a set of $n$ points $\ve{x}_q\in\R^p$, with $q=1,\ldots,n$, each of them living in the $p$-dimension space of the features. A \emph{feature} is a distinctive attribute of each element of the data set. Each point $\ve{x}_q$ is associated with one of two different classes with binary labels $y_q\in\{-1,1\}$, with $q=1,\ldots,n$, depending on the specific classification problem that we are solving.

\subsection*{\acl{svm}}

\acf{svm}\,\cite{HastieESL2009} is one of the first \ac{ml}-model originally used to carry out classification tasks. 
The \ac{svm} training consists in finding the hyperplane that separates the elements $\ve{x}_q$ in two groups: one with the label $y_q=1$ and the other with $y_q=-1$. The final hyperplane, solution of the classification, is the one having the maximum geometrical distance from the two parallel hyperplanes that are defined by the subsets of $\ve{x}_q$ called the \emph{support} sets. When the data is not linearly separable, the \emph{kernel trick} allows to increase the dimension of the features space in a way that the data becomes linearly separable in the new space.

Historically, \ac{svm} is a generalisation of the \ac{svc} that, in turn, is an improved version of the \ac{mmc} \cite{HastieESL2009}. 
\acp{mmc} aim at finding the hyperplane separating the two aforementioned classes of points, such that the distance between the hyperplane and the nearest points of the classes (commonly denoted as \emph{margin}) is maximised. If the points of the data set are not linearly separable, then the value of the margin is negative. In such a case, the \acp{mmc} cannot be adopted. \acp{svc} increase the performance of \acp{mmc}, by allowing some points of the data set, called \emph{slack variables}, to be in the opposite part of the hyperplane with respect to the others of the belonging class. If the data set exhibits a non-linear bound between the two classes of points, \acp{svc} are not able to correctly separate them, albeit the method returns a solution. Finally, \acp{svm} extend the capabilities of \acp{svc} by increasing the number of dimensions of the feature-space, such that in the new space the data set becomes linearly separable.

\subsection*{\acl{mlp}}\label{sec:mlp}

There are several classification problems (as for example the ImageNet Large Scale Visual Recognition Challenge\,\cite{russakovsky2015imagenet} employing millions of images with hundreds of categories) that are solved through \ac{svm} but with a quite high residual classification error. For this reason, in order to improve the performance in solving classification problems, \acp{ann} have been recently (re-)introduced as more-performing tools, and since 2012 have been extensively used\,\cite{russakovsky2015imagenet,schmidhuber2015deep,krizhevsky2017imagenet,emmert2020introductory}.

A \ac{mlp} is composed of a variable number of fully connected layers, each of them with a variable number of artificial neurons. A single artificial neuron with $I$ inputs ($\ve{x}$) calculates the output as
\begin{equation*}
\hat{y} \equiv \sigma(\ve{w}^T\cdot\ve{x}+b)
\end{equation*}
that is the weighted sum of the inputs $\ve{x}\in\R^I$ with weights $\ve{w}\in\R^I$, plus a bias term $b\in\R$, followed by a nonlinear activation function $\sigma:\R\rightarrow\R$. The most common activation functions $\sigma(\cdot)$ are: The \emph{sigmoid} $\sigma(x) \equiv (1+e^{-x})^{-1}$; the \emph{hyperbolic tangent} $\sigma(x) \equiv \tanh(x)$; and the \emph{rectifier} $\sigma(x) \equiv \max(0,x)$\,\cite{glorot2011deep,nair2010rectified}. A single \ac{mlp}-layer, composed of $O$ neurons with $I$ inputs, calculates 
\begin{equation*}
\hat{\ve{y}} \equiv \sigma(W^T\cdot\ve{x}+\ve{b})\,,
\end{equation*}
where $\hat{\ve{y}}\in\R^O$ is the output vector, $W\in\R^{I\times O}$ is a matrix that collects all the weight vectors of the single neurons, and $\ve{b}\in\R^O$ is the vector of the biases. Finally, an \ac{mlp} with $L$ layers calculates
\begin{equation}
    \ve{h}[l] \equiv \sigma\left(W[l]^T\cdot\ve{h}[l-1]+\ve{b}[l]\right)
\end{equation}
with $l=1,\ldots,L$ (index over the layers) and $\ve{h}[0] \equiv \ve{x}$. Thus, $\hat{\ve{y}} \equiv \ve{h}[L]$ is the output of the \ac{mlp}, where $W[l]$ and $\ve{b}[l]$ are, respectively, the weights and the biases of the $l$-th layer. Also the activation function may change depending on the specific layer. More concisely, the \ac{mlp} can be denoted by the function
\begin{equation}\label{eq:mlp}
    \hat{\ve{y}} = f(\ve{x};\theta,\xi)
\end{equation}
of the inputs $\ve{x}$. The function $f$ is parametrised by the set $\theta \equiv \{W[1],\ve{b}[1],\dots,W[L], \ve{b}[L]\}$ and by the fixed hyperparameters $\xi$ defining the number, the dimension, and the activation functions of the \ac{mlp} layers.

\subsection*{Supervised Training}

Let us now introduce the supervised learning process. For the sake of clarity, we just refer to the training of the \ac{mlp}; however, the same notions can be applied in general to the supervised learning of vast majority of \acp{ann}.

Eq.\,(\ref{eq:mlp}) behaves like a generic function approximator\,\cite{hornik1989multilayer}. Ideally, in the training process we would like to find the parameters
\begin{equation}\label{eq:argmin}
\theta^* = \argmin_\theta L_\mathcal{D}(\theta,\xi)
\end{equation}
that minimise the theoretical risk function
\begin{equation}\label{eq:risk_function}
    L_\mathcal{D}(\theta,\xi) \equiv \E_{(\ve{x},\ve{y})\sim\mathcal{D}}\left[\ell\left(f(\ve{x};\theta,\xi),\ve{y} \right)\right],
\end{equation}
i.e., the expected value of $\ell$ for $(\ve{x},\ve{y})$ sampled from the distribution $\mathcal{D}$ that generates the data set\,\cite{ShaiUML2014}. 
In \Cref{eq:risk_function}, $\ell:\R^{O\times O}\rightarrow\R^+$ denotes the loss function (usually taken as a differentiable function, apart removable discontinuities) that measures the distance between the prediction $\hat{\ve{y}}$ and the desired output $\ve{y}$. In general, the distribution $\mathcal{D}$ is unknown; thus, the minimisation problem in \Cref{eq:argmin} cannot be neither calculated nor solved. Indeed, one can dispose of a finite set $S=\{(\ve{x},\ve{y})_1,\dots,(\ve{x},\ve{y})_n\}$ of samples, to train, validate and test the ML-model. 
By considering the partition $\{S_{tr},S_{va},S_{te}\}$ of $S$, the theoretical risk function is approximated by the empirical risk function
\begin{equation}\label{eq:optimization}
    L_{S_{tr}}(\theta,\xi) = \frac{1}{|S_{tr}|}\sum_{(\ve{x},\ve{y})\in S_{tr}}\ell\left(f(\ve{x};\theta,\xi),\ve{y} \right)
\end{equation}
that is the arithmetic mean of the loss function $\ell$ evaluated on all the samples of the training set $S_{tr}$\,\cite{ShaiUML2014}. By minimising the empirical risk function $L_{S_{tr}}(\theta,\xi)$ with respect to $\theta$, the \ac{mlp} is trained and $\theta^{*}$ obtained. Then, the validation set $S_{va}$ is used to compute the empirical risk $L_{S_{va}}(\theta^{*},\xi)$ that takes as input the optimal parameters attained by the minimisation of $L_{S_{tr}}$ (training stage). This procedure allows to check if the ML-model works also for unseen data. Notice that the minimisation of the training risk function $L_{S_{tr}}(\theta,\xi)$ with respect to $\theta$ is performed step-by-step over time. After each step (also called \emph{epoch}), the validation risk $L_{S_{va}}(\theta^{*},\xi)$ is evaluated, and the minimisation procedure is stopped when the time-derivative of $L_{S_{va}}(\theta^{*},\xi)$ becomes positive for several epochs, thus showing \emph{overfitting}\,\cite{caruana2001overfitting}. In case such time-derivative remains negative or constant over time, the procedure is ended after a predefined number of epochs. The validation set $S_{va}$ can be also used to explore other configurations $\xi$ of the ML-model: this process is called \emph{hyperparameters optimization}. In particular, after completing the training procedure using two different set of hyperparameters $\xi$ and $\xi'$, we obtain two minima $\theta^{*}$ and $\theta'^{*}$, and then compare $L_{S_{va}}(\theta^{*},\xi)$ with $L_{S_{va}}(\theta'^{*},\xi')$ to also choose the best hyperparameter. Finally, we use the test set $S_{te}$ to calculate a significant metric (in our case, the classification accuracy) and report the results.

Regarding the hyperparameters optimization, it can be performed in different ways. The most basic technique is called \emph{grid search} whereby the training and validation are carried out on a specific set of hyperparameters configurations. The \emph{random grid search} considers configurations where each hyperparameter is randomly chosen within an a-priori fixed range of values. It has been proved to be more efficient than standard grid search\,\cite{bergstra2012random}. A more sophisticated class of optimization methods is the \emph{Bayesian optimization}\,\cite{snoek2012practical} that updates, after the training of each hyperparameters configuration, a Bayesian model of the validation error. The best hyperparameters configuration is thus chosen as the one allowing for the lower guess validation error.

\subsection*{Minimisation algorithms}

The most used optimisation algorithm to minimise \Cref{eq:optimization} is the \ac{sgd}\,\cite{bottou2010large,zhang2004solving,bottou2018optimization} and its adaptive variants, such as \ac{adam}\,\cite{kingma2014adam}, that changes the value of the learning rate $\eta$ (i.e., the descent step) at each iteration. After having calculated the predictions $\hat{\ve{y}}$, the loss function $\ell(\hat{\ve{y}},\ve{y})$ is propagated backwards (\emph{backpropagation}) in the \ac{ann} and its gradient in the weight space is calculated. Overall, the optimisation process consists in iteratively updating the value of the weights $\theta$ according to the relation
\begin{equation*}
    \theta_i=\theta_{i-1}-\eta\nabla_{\theta} L_{S_b}(\theta_{i-1},\xi),
\end{equation*}
where $i$ is the index for the descent step and $S_b\subset S_{tr}$ denotes the $b$-th set of samples, taken from the training set and used for the computation of the gradient. If $S_b\equiv S_{tr}$, the algorithm is called \emph{batch} \ac{sgd}; if $S_b$ contains only one element is called \emph{on-line} \ac{sgd}; finally, the most common approach (we use it here) is \emph{mini-batch} \ac{sgd} that consider $|S_b|=B$ with $B$ a fixed dimension\,\cite{bottou2018optimization}. Hence, the update of $\theta$ follows the descent direction of the gradient, with a magnitude determined by the learning rate $\eta$. 

Now, let us introduce the specific loss function $\ell$ considered in this paper. For classification problems with two or more classes, a common choice for $\ell$ is the \emph{categorical cross entropy}, which is defined as
\begin{equation}
    \ell(\hat{\ve{y}}, \ve{y})=-\sum_{j=1}^{O} y^{(j)}\log\hat{y}^{(j)}.
\end{equation}
This function measures the dissimilarity between two or more probability distributions. Thus, to properly use the categorical cross entropy, it is convenient to choose the desired outputs $\ve{y}$ as Kronecker delta functions centered around the indices associated with each class to be classified. The model output $\hat{\ve{y}}$, instead, is normalised so that it represents a discrete probability distribution, i.e., a vector of positive elements summing to $1$. This operation is obtained by using \emph{softmax}\,\cite{GoodfellowDL2016} as the activation function of the last layer:
\begin{equation}
    \sigma^{(i)}(\ve{z}) \equiv \frac{e^{z^{(i)}}}{\displaystyle{\sum_{j=1}^O e^{z^{(j)}}}}
\end{equation}
where $\sigma(\ve{z})$ is the vector having as elements $\sigma^{(i)}(\ve{z})$, with $i=1\dots,O$, and $\ve{z}$ denotes the output of the last layer before the activation function.

In the experiments, the activation functions for the hidden layers of the \ac{mlp} have been chosen among the sigmoid, hyperbolic tangent and rectifier functions accordingly to the hyperparameters optimization.

\subsection*{\aclp{rnn}}

A Recurrent Neural Network (RNN) is an \ac{ann} specialised for sequence processing when the data set is expressed as
\begin{equation}\label{eq:S_RNN}
    S=\left\{\left(\left\{\ve{x}_1,\dots,\ve{x}_{\tau_1}\right\},\ve{y}\right)_1,\dots,
    \left(\left\{\ve{x}_1,\dots,\ve{x}_{\tau_n}\right\},\ve{y}\right)_n\right\},
\end{equation}
where $\tau_{r}$ defines the number of elements of the $r$-th sequence. \acp{rnn} can be used in tasks regarding \ac{nlp}\,\cite{goldberg2017neural,martina2020classification,MartinaThesis,tang2015document}, time series analysis\,\cite{jaeger2004harnessing} and, in general, all the tasks involving ordered set of data\,\cite{lipton2015critical}. Note that, in general, for \emph{sequence-to-sequence} problems also $\ve{y}$ can be a sequence of elements, as for example in machine translation where the inputs and outputs of the \ac{rnn} are sentences in different languages\,\cite{sutskever2014sequence}. In this paper, sequence-to-sequence problems are not considered, and we thus consider $\tau_1\equiv\dots\equiv\tau_n \equiv \tau$.

\begin{figure}[!ht]
  \centering
    \includegraphics[width=0.8\textwidth]{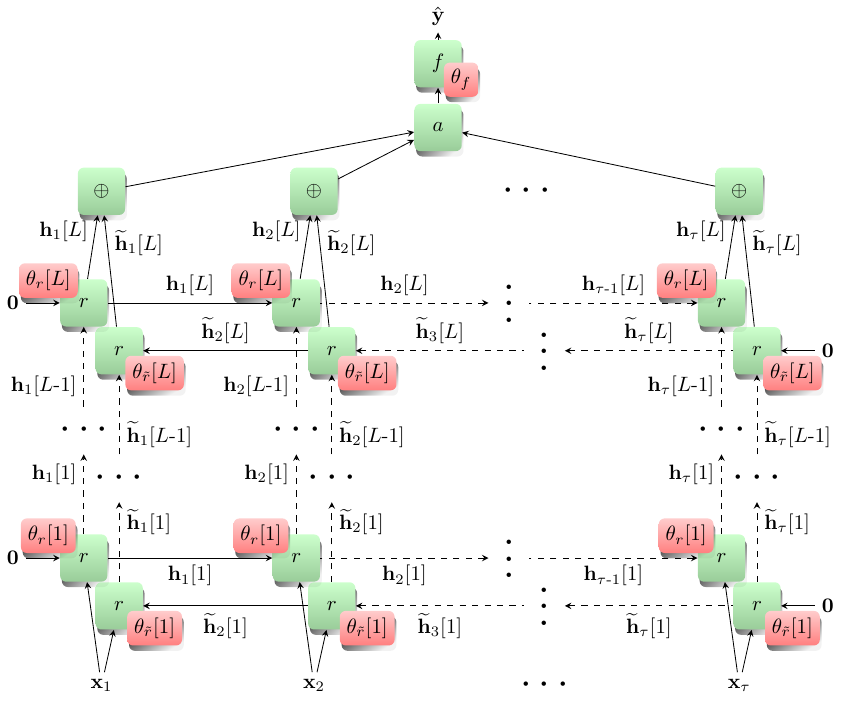}
  \caption{Diagram of a bidirectional multi-layer \ac{rnn} where the nonlinear function $r$ is defined in \Cref{eq:rnnF,eq:rnnB}, $a$ can be defined either with \Cref{eq:aggAtt} or \Cref{eq:aggMax}, $f$ is provided by \Cref{eq:rnnAll}, and $\oplus$ denotes concatenation. The input sequence $\ve{x}_t$, with $t=1,\dots,\tau$, is processed sequentially in both directions by the function $r$ that is parametrised by the shared sets of weights $\theta_r[1]$ and $\theta_{\Tilde{r}}[1]$ for the forward and backward directions, respectively. The hidden representations $\ve{h}_t[1]$ and $\widetilde{\ve{h}}_t[1]$, in turn, are processed by the subsequent layers, parametrised by a different sets of weights, so as to obtain the final hidden representations $\ve{h}_t[L]$ and $\widetilde{\ve{h}}_t[L]$. Finally, $a$ performs the aggregation of the last hidden representations adopting the attention mechanism (\ref{eq:aggAtt}) or the max pooling in \Cref{eq:aggMax}. The classification is performed by the function $f$ that is parametrised by $\theta_f$. The simpler form of aggregation in \Cref{eq:aggregation} is not depicted in the figure.}
  \label{fig:rnn}
\end{figure}

A \ac{rnn} is defined by the \emph{recurrent} relation
\begin{equation}\label{eq:recurrentRelation}
    \ve{h}_t=r\left(\ve{x}_t,\ve{h}_{t-1};\theta_r,\xi\right)
\end{equation}
where $t\in\{1,\dots,\tau\}$, $\ve{h}_t\in\R^d$ is a $d$-dimensional vector with $d$ being an hyperparameter belonging to $\xi$ and  $\ve{h}_0=\ve{0}$ (vector of zeros). The recurrent relation (\ref{eq:recurrentRelation}) defines $\tau$ hidden representations $\ve{h}_t$ (to be seen as a \emph{memory}) of the input sequence $\{\ve{x}_1,\dots,\ve{x}_{\tau}\}_q$ with $q=1,\ldots,n$. If the function $r$ is implemented as an \ac{mlp} (\ref{eq:mlp}) that takes as input the concatenation $\ve{x}_t\oplus \ve{h}_{t-1}$ (usually called ``vanilla \ac{rnn}''), the model suffers the so-called \emph{vanishing gradient problem}\,\cite{hochreiter1998vanishing,bengio1994learning} such that the weights of the last layers of the \ac{rnn} are updated only with respect to the more recent input data. The vanishing gradient problem occurs when the backpropagation is performed on a high number of layers, as it could happen in our case with a large value of $\tau$ (thus meaning long input sequences). In this regard, to mitigate the vanishing gradient problem, \ac{lstm}\,\cite{hochreiter1997long} and \ac{gru}\,\cite{cho2014learning} have been introduced. These methods use learned gated mechanisms, based on current input data and previous hidden representations, to control how to update the current hidden representation $\ve{h}_t$. Specifically, if \ac{lstm} is used, \Cref{eq:recurrentRelation} needs to be slightly modified as
\begin{align}
    \ve{s}_t &= v\left(\ve{x}_t,\ve{s}_{t-1};\theta_{v},\xi\right)\label{eq:lstm_eqs}\\
    \ve{h}_t &= r\left(\ve{x}_t,\ve{s}_t,\ve{h}_{t-1};\theta_{r},\xi\right)
\end{align}
where $\ve{s}_0 = \ve{0}$ and $v$, $r$ are, as usual, nonlinear functions. Both for \ac{gru} and \ac{lstm}, the nonlinearity of the recurrent relations is due to the adoption of the hyperbolic tangent and sigmoid functions, where the latter are employed only for the gating mechanism. It is worth noting that in \Cref{eq:lstm_eqs} $\ve{s}_t$ is a state vector that allows to differently propagate over time specific elements of the hidden representations $\ve{h}_t$ depending on the input data. This means that, at any time $t$, the hidden representation $\ve{h}_t$ depends not only on the input $\ve{x}_t$ and the previous hidden representation $\ve{h}_{t-1}$ but also on the state vector $\ve{s}_t$. For further details, refer to Refs.\,\cite{hochreiter1997long,cho2014learning,goldberg2017neural}.

\acp{rnn} are usually considered deep-learning models, due to the high number of layers, when they are unfolded on the sequence dimension for $t=1,\dots,\tau$. The key aspect of deep learning is the automatic extraction of features by means of the composition of a large number of layers; an increasing (deep) number of layers is typically used to extract features with increasing complexity\,\cite{lecun2015deep}.

Moreover, \acp{rnn} can be extended considering more layers\,\cite{graves2013speech} and processing the data bidirectionally\,\cite{schuster1997bidirectional}. Regarding the latter, one can define two sets of hidden representations: One for the forward and the other for the backward direction, where the $t$-th hidden representation depends respectively on the $(t-1)$-th or $(t+1)$-th one. More formally,
\begin{align}
    \ve{h}_t[l]&=r\left(\ve{h}_t[l-1],\ve{h}_{t-1}[l];\theta_r[l],\xi\right) \label{eq:rnnF} \\
    \widetilde{\ve{h}}_t[l]&=r\left(\widetilde{\ve{h}}_t[l-1],\widetilde{\ve{h}}_{t+1}[l];\theta_{\Tilde{r}}[l],\xi\right) \label{eq:rnnB}
\end{align}
with $l=1,\dots,L$ and $\ve{h}_t[0]\equiv\widetilde{\ve{h}}_t[0]\equiv\ve{x}_t$. 

\subsection*{Classification with \acp{rnn}}

Now, let us explain how to use the hidden representations to calculate the prediction $\hat{\ve{y}}$ in output from the ML-model. The common approach to calculate the prediction in classification problems is to use the \ac{rnn} as an encoder of the sequence and to scale the dimension of the last hidden representation $\ve{h}_{\tau}[L]\oplus\widetilde{\ve{h}}_1[L]$ (in the more general case of bidirectional models) to the one of the output vector. This scaling can be done through a fully connected layer, or, more in general, by means of an \ac{mlp}, i.e.,
\begin{align}
    \ve{a}&\equiv\ve{h}_{\tau}[L]\oplus\widetilde{\ve{h}}_1[L]\label{eq:aggregation}\\
    \hat{\ve{y}}&=f\left(\ve{a};\,\theta_f,\xi\right).\label{eq:rnnAll}
\end{align}
Then, we can use \ac{sgd} to minimise an empirical risk function similar to \Cref{eq:optimization} of \acp{mlp}. 

It is possible to consider different forms of aggregation $\ve{a}$ for the hidden representations $\ve{h}_t[L]$, with $t=1,\dots,\tau$, instead of using only the last hidden representation as in \Cref{eq:aggregation}. In this regard, \emph{attention mechanisms}\,\cite{bahdanau2015neural,luong2015effective,chorowski2015attention}, also in hierarchical forms\,\cite{yang2016hierarchical}, perform a weighted average of the $\ve{h}_t[L]$ where the weights are learned together with the ML-model. In detail, \Cref{eq:aggregation} becomes:
\begin{align}
    \ve{u}_t&=\ve{h}_t[L]\oplus\widetilde{\ve{h}}_t[L]\label{eq:ht}\\
    \ve{v}_t&=\tanh\left(\ve{W}^T\cdot\ve{u}_t+\ve{b}\right)\label{eq:att1}\\
    \alpha_t&\equiv\frac{e^{\langle\ve{v}_t,\ve{c}\rangle}}{\displaystyle{\sum_{j=1}^{\tau}e^{\langle\ve{v}_j,\ve{c}\rangle}}}\label{eq:att2}\\
    \ve{a}&\equiv\sum_{t=1}^\tau\alpha_t\ve{u}_t,\label{eq:aggAtt}
\end{align}
where $\langle\cdot,\cdot\rangle$ denotes the dot product and $\ve{c}$ is a learned vector 
that is randomly initialised and jointly learned during the training process as in Refs.\,\cite{bahdanau2015neural,luong2015effective,chorowski2015attention,yang2016hierarchical}. Another form of aggregation $\ve{a}$ is the \emph{max pooling aggregation}, whereby each element $\ve{a}^{(j)}$ of $\ve{a}$ just refers to a single value of $t$. In this case, \Cref{eq:aggregation} equals to
\begin{equation}\label{eq:aggMax}
    \ve{a}^{(j)}=\max_t\ve{u}_t^{(j)}.
\end{equation}
where the expression of $\ve{u}_t$ is provided by \Cref{eq:ht}. In this way, each element $\ve{u}_t^{(j)}$ of the hidden representations (for $t=1,\dots,\tau$) learns to detect specific features of the input data within all the interval $[1,\tau]$.

Finally, another approach, which we do not use here, is to consider the \ac{rnn} as a transducer that produces an output sequence $\hat{\ve{y}}_t$ for $t=1,\dots,\widetilde{\tau}$ (generally $\widetilde{\tau} \neq \tau$) in correspondence of the input sequence $\ve{x}_t$ with $t=1,\dots,\tau$\,\cite{goldberg2017neural,sutskever2014sequence,graves2012sequence}.

\subsection*{Implementation of the machine learning algorithms}

All the ML-models are realized in \emph{PyTorch} and have been trained on the six different data sets 
using a DELL\textsuperscript{\textregistered} Precision Tower workstation with one NVIDIA\textsuperscript{\textregistered} TITAN RTX\textsuperscript{\textregistered} GPU with 10 Gb of memory, 88 cores Intel\textsuperscript{\textregistered} Xeon\textsuperscript{\textregistered} CPU E5-2699 v4 at 2.20GHz and 94 Gb of RAM. 

We train the \ac{ann} models in mini-batches of dimension $16$ by means of the \ac{sgd} using \ac{adam}\,\cite{kingma2014adam} and learning rate $\eta=10^{-3}$. We optimize the hyperparameters $\xi$ with \emph{ASHA}\,\cite{li2020system} as scheduler and \emph{Hyperopt}\,\cite{bergstra2011algorithms,bergstra2013making} (Hyperopt belongs to the family of Bayesian optimization algorithms) as search algorithm in the framework \emph{Ray Tune}\,\cite{liaw2018tune}. For the \ac{mlp} models, the hyperparameters optimization defines: (i) the activation functions to be used, (ii) the number of layers, and (iii) their dimension, within the following search space: $\sigma\in\{\text{relu},\text{sigmoid},\text{tanh}\}$, $L\in\{2,3,4,5,6\}$ and $\dim(\ve{h}[1])\equiv\dots\equiv\dim(\ve{h}[L])\in\{d\in\N\,|\,1\leq d\leq 512\}$. Instead, for the \ac{rnn} models the search space is $L\in\{1,2,3,4\}$ for the number of recurrent layers ($L\in\{1,2,3,4,5,6\}$ for the \nm{} task with $t_{15}=0.1$), and $\dim(\ve{h}_t[1])\equiv\dim(\ve{h}_t[1])\equiv\dots\equiv \dim(\widetilde{\ve{h}}_t[L])\equiv\dim(\widetilde{\ve{h}}_t[L])\in\{d\in\N\,|\,1\leq d\leq 512\}$ for the layers dimension. Regarding the ML-models \bigrua{} and \bilstma{}, the search space includes also the dimension of the attention layer as in \Cref{eq:att1,eq:att2}, i.e., $\dim(\ve{c})\equiv\dim(\ve{v}_1)\equiv\dots\equiv\dim(\ve{v}_\tau)\in\{d\in\N\,|\,1\leq d\leq 512\}$. In the hyperparameters optimization of all the \ac{mlp} and the \ac{rnn} models, we have also used regularization methods as \emph{weight decay}\,\cite{krogh1992simple} and \emph{dropout}\,\cite{srivastava2014dropout}. They are able to mitigate overfitting; in particular, the former adds a penalty (chosen among $\{0,10^{-4},10^{-3}\}$) to the risk function $L_S(\theta,\xi)$ with the aim to discourage large weights. Instead, using dropout, the outputs of the artificial neurons during the training are forced to zero with a probability among $\{0,0.2,0.5\}$.

\bibliographystyle{unsrt}
\bibliography{biblio}

\end{document}